\documentstyle[aps,twocolumn,psfig]{revtex}

\begin{document}
%\draft

\twocolumn[\hsize\textwidth\columnwidth\hsize\csname @twocolumnfalse\endcsname

\title{Exact single spin flip for the Hubbard model in $d=\infty$}

\author{G\"otz S.~Uhrig}

\address{Laboratoire de Physique des Solides,
Universit\'e Paris-Sud, b\^at.~510,
F-91405 Orsay}

\date{\today}

\maketitle

\begin{abstract}
It is shown that the dynamics of a single $\downarrow$-electron
interacting with a band of $\uparrow$-electrons can be calculated
exactly in the limit of infinite dimension.
 The corresponding
Green function is determined as a continued fraction. 
It is used to
investigate the stability of saturated ferromagnetism
and the nature of the ground state for two generic non-bipartite
infinite dimensional lattices. Non Fermi liquid
behavior is found. For certain dopings
 the $\downarrow$-electron is bound to the $\uparrow$-holes.\\

\\
\end{abstract}
]

%introduction

In the last years the limit of infinite dimensions $d\to \infty$,
 introduced by Metzner and Vollhardt for fermionic lattice
models\cite{metzn89a},
 has proved to be very useful in the investigation of
strongly correlated electron systems \cite{mulle89a,vollh93}.
Its merit is an important simplification reducing
lattice problems to effective one-site problems
\cite{janis91} supplemented
by a self-consistency condition \cite{georg92ajarre92}.
The limit $d\to \infty$
generates a mean-field theory
which constitutes a very successfull local
 approximation in $d<\infty$ \cite{prusc96georg96}.
The mean-field keeps its frequence dependence (dynamic MFT)
for local interactions \cite{janis92ajanis92b}.
Thus important physics is retained. The
effective one-site problem remains generally difficult
to solve\cite{prusc96georg96}.
So far, the metal-insulator
transition \cite{prusc96georg96} and
various ordering phenomena
\cite{donge91donge94bdonge95,uhrig93buhrig95d,freer93,freer95,rozen95}
have been treated successfully.
Recently, the investigation of polarized situations 
has been started \cite{held95,ulmke95}.

Here the Green function of a $\downarrow$-e$^-$ interacting
with the $\uparrow$-e$^-$ of filling $n$ (complete polarization)
is calculated exactly.
Two points will be addressed: 
(a) the stability of the Nagaoka state toward
a spin flip; (b) the nature of the ground state.

(a): The understanding of ferromagnetism is an old problem
 at the origin of the Hubbard model \cite{hubba63kanam63gutzw63}.
Saturated ferromagnetism (Nagaoka state) was proved for $U=\infty$
by Nagaoka for one e$^-$ above $n=1$ \cite{nagao66}.
Ferromagnetism has been proved recently for a number of special cases:
either the filling is such that the Fermi energy lies in a flat band
\cite{mielk91amielk91btasak92mielk93}
or the situation corresponds to half-filling of one (sub)band
\cite{lieb89,strac93strac94strac95,tasak94tasak95} or the limit
$n\to 0$ for a Hubbard chain with two
minima in the dispersion \cite{mulle95} is considered.
In $U$-$\delta$ diagrams (cf.\ fig.\ 1, 2), 
saturated ferromagnetism has been established by these results
on vertical lines.

The range of stability of the Nagaoka state has been investigated
extensively by variational methods
%\cite{roth67roth69richm69shast90basil90linde91,
%fazek90,hanis93,wurth95,hanis95}.
see e.g.\ \cite{fazek90,hanis93,wurth95,hanis95}.
More and more sophisticated trial wave functions for a single flipped
spin were used. Thereby, the region of possible local stability of the
Nagaoka state was reduced considerably for various lattices.
One has to rely so far on exact diagonalization data for
small systems to estimate how far the variational phase boundaries
are situated from the true ones; e.g.\ numerical
results for the square lattice 
propose a doping above which the Nagaoka state is unstable at
$U=\infty$ of $\delta_c=0.195$ \cite{hirsc94} or $\delta_c=0.22$
\cite{liang94} and the best rigorous upper bound is $\delta_c=0.251$
\cite{wurth96} at present. 

The present work establishes exactly
(in)stability of saturated ferromagnetism toward 
single spin flip for lattices in $d=\infty$ for all values of $\delta$.
This allows to evaluate the quality of the variational approaches
for these lattices. 

(b): For certain values of $\delta$ the $\downarrow$-e$^-$
forms bound states with the $\uparrow$-holes which 
implies a non-Fermi liquid behavior. The mobility of the 
$\downarrow$-e$^-$ is reduced decisively: it does no longer
diverge as it does in a Fermi liquid. The picture of bound
particle-hole pairs is a possible interpretation of the
metal-insulator transition occurring at $n=1$ \cite{donge94a}.
 
Consider the local Green function
for one $\downarrow$-e$^-$ $G(\omega):= G_{i,i;\downarrow}(\omega)$;
$G_{i,i;\downarrow}$ is the Fourier transform of
$-i\langle [\hat a^{\phantom\dagger}_{i;\downarrow}(t),
\hat a^\dagger_{i;\downarrow}(0) ]{} \rangle \Theta(t)$
 ($\hbar$ is set to 1; $i$ can be any site).
This $\downarrow$-e$^-$ interacts in a Hubbard model
with the Fermi sea of $\uparrow$-e$^-$.
The expectation value is taken for the Fermi sea of $\uparrow$-e$^-$.
Introducing the Liouville operator
$L:=[H,\ ]$ \cite{fuldec} $G(\omega)$ can be expressed as
$G(\omega) = \langle \hat a^{\phantom\dagger}_{i;\downarrow} (\omega-L)^{-1}
 \hat a^\dagger_{i;\downarrow} \rangle$.
The Mori/Zwanzig projection method provides $G(\omega)$ as continued
fraction (CF) with the static correlations of the Fermi sea
as coefficients \cite{fuldec}.
So, $G(\omega)$ is {\em in principle} known.
Yet, due to the rapidly increasing complexity of the coefficients with
increasing order the CF is {\em not} tractable in $d<\infty$.

For $d\to\infty$, however, the self-energy is local \cite{mulle89a}.
It is sufficient to consider
$\Sigma(\omega) := \Sigma_{i,i;\downarrow}(\omega)$.
The skeleton diagrams of $\Sigma(\omega)$
comprise {\em only} the local vertex $i$ \cite{mulle89a}. This
implies that $\Sigma(\omega)$ is the same as the self-energy
for the first site of
an effective half-infinite chain of e$^-$ which interact
only at the first site \cite{vollh93,prusc96georg96}
if their full local propagators are the same.
This is the consistency condition:
the chain Green function of the
 first site $g_\lambda(\omega)$ and the 
local lattice Green function are equal:
$G_{i,i;\lambda}(\omega)=g_\lambda(\omega),
 \lambda\in\{\uparrow,\downarrow\}$ \cite{georg92ajarre92}.
The matrix elements of the  chain 
have to be determined such that the consistency
is fulfilled. So, we are facing
two nested problems: solving the consistency condition to
find the matrix elements of the chain, and
solving the one-site problem to find the self-energy.

For a single $\downarrow$-e$^-$ the two problems
can be solved by iterating a straightforward
procedure. First, we state that the matrix elements of the chain 
describing the $\uparrow$-e$^-$ are those of the non-interacting
problem since the $\uparrow$-e$^-$ do not interact
with themselves (Recall that the ground state considered is the Fermi sea
of $\uparrow$-e$^-$). 
So these matrix elements, i.e.\ local energies and
nearest neighbor hopping elements, are those of the CF
$(a_i,b_i)$ of the non-interacting lattice DOS $\rho(\omega)$
as weight function \cite{petti85}, i.e.\ the CF of 
$G^0(\omega)$, the non-interacting local propagator.

Second, we know from Dyson's equation that
$G(\omega)=G^0(\omega-\Sigma(\omega))$ for the lattice,
and $g(\omega) = g^0(\omega)/\left(1-g^0 \Sigma(\omega) \right)$
for the chain ($\downarrow$ spin index omitted).
The identity of the local lattice proper self-energy and of
the equivalent chain quantity is used. Hence the consistency condition
requires $G^0(\omega-\Sigma(\omega)) = g^0(\omega)
/\left(1-g^0 \Sigma(\omega) \right)$.
Knowing the first CF coefficient pair $(a_0,b_0)$
of $G^0(\omega)$, i.e.\ 
$G^0(\omega)=(\omega-a_0-b_0^2 \tilde G^0(\omega))^{-1}$,
the consistency is equivalent to
$g^0(\omega)=(\omega-a_0-b_0^2 \tilde g^0(\omega))^{-1}$
and
\begin{equation}
\label{consist2}
\tilde g^0(\omega) = \tilde G^0(\omega-\Sigma(\omega))\ .
\end{equation}
By (\ref{consist2}) further
CF coefficients $(x_i,y_i), i>0$,  of $g^0$ can be calculated 
from  the CF coefficients $(s_i,t_i)$ of $\Sigma$ by
tridiagonalization. Note to this end that
(\ref{consist2}) can be viewed as
the Green function belonging to the one-particle Hamiltonian of a 
semi-infinite chain (CF of $\tilde G^0$) with the replicas of the 
chain of the CF of $\Sigma$ attached to each site.
The {\em crucial} point is that 
$(s_0,t_0) \ldots (s_n,t_n)$ are sufficient to calculate 
$(x_0,y_0) \ldots (x_{n+1},y_{n+1})$.

Third, for the calculation of
$\Sigma(\omega)$ the projection method \cite{fuldec} provides
\begin{equation}
\label{sigma1}
\Sigma(\omega) = U n + U^2\left\langle 0\left|\hat n_{0;\uparrow} Q
(\omega-QLQ)^{-1} Q \hat n_{0;\uparrow}\right|0\right\rangle
\end{equation}
where $|0\rangle$ stands for the Fermi sea of $\uparrow$-e$^-$
plus the $\downarrow$-e$^-$ at the first site of the chain;
 $\hat n_{0;\uparrow}:= \hat c_{0;\uparrow}^\dagger
\hat c^{\phantom\dagger}_{0;\uparrow}$;
 $Q$ projects out states without any particle-hole
excitation, $U$ is the interaction, and $L$ is the commutation with 
\begin{eqnarray}\nonumber
\hat H_{\rm\scriptstyle chain} &=& U \hat n_{0;\uparrow} \hat n_{0;\downarrow}
+\sum_{i=0}^\infty a_i \hat c_{i;\uparrow}^\dagger
\hat c^{\phantom\dagger}_{i;\uparrow} + 
b_i (\hat c_{i+1;\uparrow}^\dagger
\hat c^{\phantom\dagger}_{i;\uparrow} +  \mbox{h.c.})
\\
&+& \sum_{i=0}^\infty x_i \hat c_{i;\downarrow}^\dagger
\hat c^{\phantom\dagger}_{i;\downarrow} + 
y_i (\hat c_{i+1;\downarrow}^\dagger
\hat c^{\phantom\dagger}_{i;\downarrow} +  \mbox{h.c.})
\ .
\end{eqnarray}

The essential difference to the problem at finite density of
$\downarrow$-e$^-$ is that for a single $\downarrow$-e$^-$ the ground
state $|0\rangle$ is known beforehand.
The CF of (\ref{sigma1}) is
found again by tridiagonalization, i.e.,
the Lanczos algorithm is used with
$\left.\left. Q \hat n_{0;\uparrow}\right|0\right\rangle$ as first
state.
The subspaces reached by the iterative application of 
$QLQ$ are spanned by states
which are products of equal numbers of $\uparrow$-creation
and $\uparrow$-annihilation operators plus one $\downarrow$-creation
operator on certain chain sites acting on the $\uparrow$-Fermi sea.
Components without any particle-hole excitation are projected out.
The {\em crucial} point is that 
$(x_0,y_0) \ldots (x_n,y_n)$ are sufficient to calculate 
$(s_0,t_0) \ldots (s_{n+1},t_{n+1})$.

Hence,
 the  nested problems of the consistency
 and of the effective one-site problem
can be solved by alternating  tridiagonalization steps
for (\ref{consist2}) with steps for (\ref{sigma1}).

Results will be presented for generalizations of the
$d=3$ fcc lattice. 
One is interested in non-bipartite, electron doped lattices
for their enhanced tendency toward ferromagnetism 
due to their asymmetric DOS with finite lower band edge
\cite{hanis95,hirsc94}. In contrast, 
 the Nagaoka state is even completely unstable  for
the hypercubic lattice in $d=\infty$ 
\cite{fazek90}.
The properties of the lattice are essential for ferromagnetism.
Finding stable ferromagnetism on non-bipartite lattices 
with local $\Sigma$ is thus no contradiction to the robustness
of local Fermi liquids \cite{engel95} in the $d=3$ continuum
limit (spherical Fermi surface corresponding to $n\to 0$
on the sc lattice). Additionally, Fermi liquid (FL) theory is not 
valid for all $\delta$ and $U$, see below and ref.\ \cite{ulmke95}.

For simplicity, we consider 
$\hat H_{\rm kin}=t\sum_{\langle i,j\rangle;\lambda}
\hat a^\dagger_{i;\lambda}
\hat a^{\phantom\dagger}_{j;\lambda}$, $\langle i,j\rangle$ nearest 
neighbors with $t>0$
and $n<1$ which is equivalent to $n>1$ and the usual $t<0$.
A large DOS at the
lower band edge is favorable for the Nagaoka state since it lowers
the kinetic energy necessary for polarization.

The half hypercubic (hh) lattice is
made of the even sites of the hypercubic (hc) lattice.
Nearest-neighbor sites are linked by $t$,
sites at distance 2 are linked by $t/2$.
Then the dispersions are related by
$\varepsilon_{\rm hh}({\bf k})= \varepsilon^2_{\rm hc}({\bf k})-1$
if $t=1/d$ on the hh and 
$t=1/\sqrt{2d}$ on the hc.
Hence
$\rho_{\rm hh}(\omega) = \rho_{\rm hc}(\sqrt{1+\omega})/(2\sqrt{1+\omega})$
\cite{polen} (recall 
$\rho_{\rm hc}(\omega) = \exp(-\omega^2/2)/\sqrt{2\pi}$ for $d\to \infty$
\cite{metzn89a}).
Omitting the $t/2$ hops leaves the DOS for $d=\infty$
essentially unchanged. Yet the dispersion
acquires a term $-t\sum_{i=1}^d \cos(2k_i)$ which leads to a foot
 extending down to $\omega=-2$. On $d\to \infty$
the weight of this foot vanishes \cite{polen}, yet the foot
is decisive for the single spin flip energy for any 
$d<\infty$.
The hh  with foot in $d=3$ is the fcc lattice and in $d=2$ a square lattice.

\begin{figure}
\setlength{\unitlength}{1cm}
\begin{picture}(8.2,6.3)(-1.2,0.7)
{\psfig{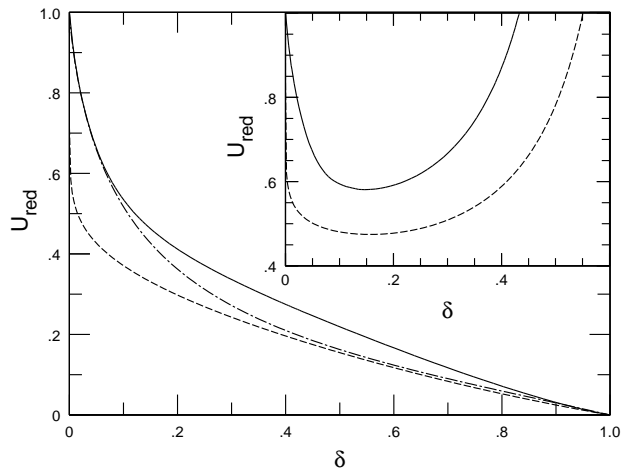}}
\end{picture}
\caption{
Half hypercubic lattice: above the solid line the Nagaoka
is stable toward spin flip.
Main figure: hh without foot; dashed-dotted:
variational result from (\ref{bound}).
Inset: hh with foot; maximum doping
$\delta_c=0.4330$; minimum interaction $U_{{\rm red}}=0.5809$;
$U_{\rm red}:=U/(U+U_{\rm BR})$ with $U_{\rm BR}=6.8588$.
}
\end{figure}
The sites of a new layer 
of the laminated (lam) lattice are above
the interstices of the layer below. The basis vectors
are defined by ${\bf b}_{n+1}=(n+1)^{-1}\sum_{i=1}^n {\bf b}_i
+\sqrt{(2+n)/(2+2n)} {\bf e}_{m+1}$
with ${\bf b}_1= {\bf e}_1$, where ${\bf e}_i$ are cubic unit
vectors \cite{mullepriv}.
On scaling $t=1/d$, one yields for $d\to \infty$ the 
exponential DOS 
$\rho_{\rm lam}(\omega) = \exp(-(\omega/2+1))\Theta(\omega+2)/2$.
The lam in $d=3$ is the fcc lattice, in $d=2$
it is the triangular lattice. Hence it is even a better
generalization of the known non-bipartite lattices than the hh.

In figs.\ 1, 2 the exact phase boundaries above which the Nagaoka state
is locally stable are shown. To compare  different
lattices the reduced interaction $U_{\rm red}:=U/(U+U_{\rm BR})$
is used where $U_{\rm BR}$ is defined from the kinetic energy
at half-filling: 
$U_{\rm BR}:=-16\int_{-\infty}^{\mu_0}\omega \rho(\omega)d\omega$.
The result (inset fig.\ 1) aggrees with the
$T\to 0$ extrapolation to saturated
magnetization of QMC data for the hh lattice 
at $\delta=0.42, U_{\rm red}=0.45198$  \cite{ulmke95}.

\begin{figure}
\setlength{\unitlength}{1cm}
\begin{picture}(8.2,6.3)(-1.2,0.7)
{\psfig{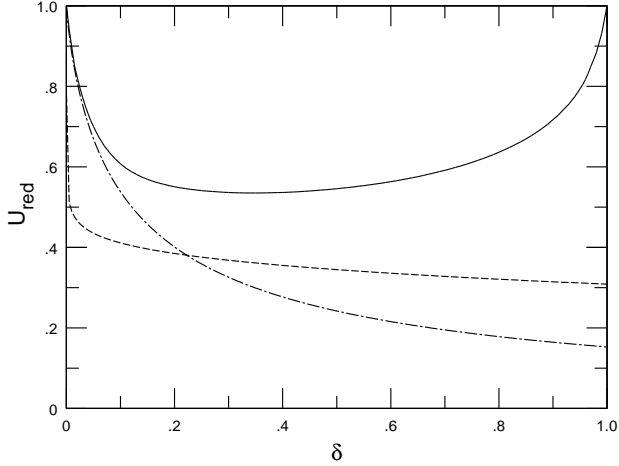}}
\end{picture}
\caption{Laminated lattice: same as in fig.\ 1;
minimum at $U_{{\rm red}}=0.5351$; $U_{\rm BR}=16\ln 2$}
\end{figure}
The transition interaction is found from the zero of the 
spin flip energy $E_{\rm SF}=\omega_0-\mu$.  Taking out
a $\uparrow$-e$^-$ gains $\mu$
and inserting a $\downarrow$-e$^-$ costs at least $\omega_0$ where
$\omega_0$ is the lower band edge of $G(\omega)$
The band edge $\omega_0$ is computed by extrapolating
the approximate band edge $\omega_{\rm a}(j)$ 
($j$ depth of the CF, here $j\leq 53$) to $j=\infty$.
The approximate values are found from 
\begin{equation}
\label{deter}
\varepsilon_{\rm b} = \omega_{\rm a}(j)-\Sigma_{\rm a}(\omega_{\rm a};j).
\end{equation}
They lie always below the band edge $\omega_\Sigma(j)$ of
$\Sigma_{\rm a}(\omega;j)$: 
Re$\, \Sigma_{\rm a}(\omega;j)$ diverges to $-\infty$ on 
$\omega\to \omega_\Sigma(j)$ because 
the spectral density of $\Sigma_{\rm a}(\omega;j)$
consists of $\delta$-peaks for $j<\infty$.
Thus (\ref{deter}) has a zero in the interval $(-\infty,\omega_\Sigma(j))$
even though the weight of lowest lying $\delta$-peak
may become arbitrarily small on $j\to\infty$.
For $j=\infty$, however, $\varepsilon_{\rm b}>\omega_0
-\Sigma(\omega_0)$ is possible since the spectral density of $\Sigma$
is generally continuous. The inequality implies that 
$G_0(\omega-\Sigma)$ acquires an imaginary part due to
the imaginary part of $\Sigma$ and {\em not} because the
lowest scattering state, renormalized by $\Sigma$, can
be excited. The latter, however, is a necessary condition for
quasi-particles to be the elementary excitations and thus
for the applicability of FL theory.

The non-solid lines are variational results.
The dashed curves result from a scattering type ansatz
\begin{equation}
\label{scatter}
\psi({\bf q}) \! = \!\!
 \sum_{j,i} \frac{e^{-i{\bf q}{\bf r}_j}}{\sqrt{N}} (
\frac{u}{N} \hat a^\dagger_{j;\downarrow}
 + u_i\, \hat a^{\phantom\dagger}_{j;\uparrow}
\hat a^\dagger_{j+i;\uparrow} \hat a^\dagger_{j;\downarrow}
)
|{\rm FS}'\rangle
\end{equation}
where $|{\rm FS}'\rangle = 
\hat a_{{\bf k}_{\rm F};\uparrow}|{\rm FS}\rangle$
(${\bf k}_{\rm F}$ Fermi wave vector, $N$ \# of sites) \cite{hanis96}.
The variational variables are $u$ and $\{u_i\}$. 
The optimum is attained for a ${\bf q}$ with 
$\varepsilon({\bf q})=\varepsilon_{\rm b}$.
The dashed-dotted lines come from an ansatz for a bound state between 
a $\downarrow$-e$^-$ and a $\uparrow$-hole
\begin{equation}
\label{bound}
\Phi({\bf q}) = \sum_{{\bf k} \in {\rm FS}} v_{{\bf k}} 
\hat a^\dagger_{{\bf k}+{\bf q};\downarrow} 
\hat a^{\phantom\dagger}_{{\bf k};\uparrow} |{\rm FS}\rangle \ .
\end{equation}
The variational variables are $\{ v_{{\bf k}}\}$.
The optimum is found for ${\bf q} \to 0$
unlike the results for the square lattice \cite{wurth95}
where ${\bf q}=(0,\pi)$ was optimal.

Comparison of the solid lines with the dashed ones in figs.\ 1, 2
reveals that an ansatz like (\ref{scatter}) gives a good
idea what the stability region looks like for different lattices.
One recognizes the influence of the singularity of $\rho(\omega)$
at $\varepsilon_{\rm b}$: an inverse square root leads to a Nagaoka state
even for $n\to 0$ and $U\to 0$, a constant implies a Nagaoka state
for $n\to 0$ but only for $U\to \infty$ (in the exact result),
 and a foot, i.e.\ a low DOS
at $\varepsilon_{\rm b}$, leads to $\delta_c<1$.
The ansatz (\ref{scatter}) 
fails especially to describe the behavior at large $U$.

Close to $\delta=0$, the dashed-dotted lines of
(\ref{bound}) obviously capture the right physics since they are
asymptotic to the exact curves. This leads to the hypothesis
that the $\downarrow$-e$^-$ is actually bound to the $\uparrow$-holes.
It is supported by the fact that for small $\delta$ the
curves for the hh with and without foot are numerically 
indistinguishable. The energy of a bound state is not changed by a
foot of infinitesimal weight whereas the energy of the
lowest scattering state is.

\begin{figure}
\setlength{\unitlength}{1cm}
\begin{picture}(8.2,6.3)(-1.2,0.7)
{\psfig{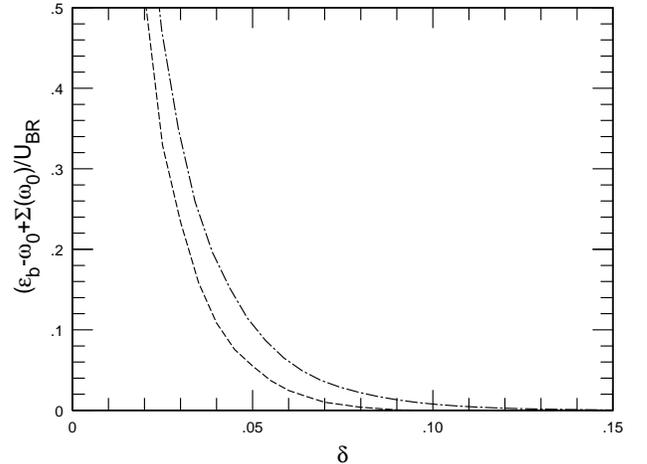}}
\end{picture}
\caption{Distance of $\omega_0-\mbox{Re}\, \Sigma(\omega_0))$ 
from the free band edge $\varepsilon_{\rm b}$ in units of $U_{{\rm BR}}$
at the phase boundary. Finite values indicate non-Fermi liquid behavior.
Dashed-dotted: hh without foot; dashed: lam.}
\end{figure}
The nature of the ground state of one $\downarrow$-e$^-$
at the phase boundary is further elucidated by fig.\ 3.
The data is obtained by extrapolating 
$\mbox{Re}\, \Sigma_{\rm a}(\omega_0;j)$ at the
lower band edge $\omega_0$ determined previously.
For larger $\delta$ the relation $\varepsilon_{\rm b}=
\omega_0-\Sigma(\omega_0)$ is found to hold which is
necessary, not sufficient, for FL behavior 
(see discussion after eq.\ (\ref{deter})).
A definite decision on the applicability of FL theory for larger 
values of $\delta$ is not yet possible,
 but see ref.\ \cite{ulmke95}.
For small $\delta$, we find definitely non-FL behavior
since  $\varepsilon_{\rm b} > \omega_0-\Sigma(\omega_0)$.
The physical interpretation suggested from the success
of the bound state ansatz (\ref{bound})
for small $\delta$ and large $U$ is that
bound $\downarrow$-e$^-$-$\uparrow$-hole states replace the quasiparticles
as elementary excitations.
From $1=\varepsilon_{\rm b}-\omega_0+\Sigma(\omega_0)$ 
(cf.\ fig.\ 3)  we find the doping
 $\delta=0.044$ below which the exact curves
for the hh with and without foot coincide in fig.\ 1.

The interpretation of $\varepsilon_{\rm b}>\omega_0+\Sigma(\omega_0)$
as indication for a  binding phenomenon is corroborated
by the resulting finite conductivity/mobility $\sigma$
of the $\downarrow$-e$^-$.  
For $T\to 0$, one has 
$\sigma\propto {\rm Im} G(\omega_0)/{\rm Im} \Sigma(\omega_0)$
\cite{uhrig95a}.
This is bounded from above according to
${\rm Im} G(\omega_0)/{\rm Im} \Sigma(\omega_0) =
\int_{\varepsilon_{\rm b}}^\infty
\rho_0(\omega)(\omega-\omega_0+\Sigma(\omega_0))^{-2}d\omega
 < (\varepsilon_{\rm b}-\omega_0+\Sigma(\omega_0))^{-2}$.
If the system displayed no binding phenomenon (e.g.\ in a FL) 
one would have a diverging $\sigma$ since ${\rm Im} \Sigma(\omega)$ vanishes
generally faster than ${\rm Im} G(\omega)$ on $\omega\to\omega_0$.

In summary, this work 
introduced a method (nested application of two
tridiagonalisations) to calculate the dynamics
of a single spin flip as continued fraction for $d\to \infty$
or, equivalently, for the local approximation.
The usual, cumbersome self-consistency condition
 \cite{vollh93,georg92ajarre92,prusc96georg96}
is solved in this case exactly.

By this method, the region of local stability
 of saturated ferromagnetism was computed exactly for two  
non-bipartite lattices and compared to variational results.
The latter yield a good qualitative 
impression of the local stability region but fail to describe
it quantitatively.

For small $\delta$ (large $U$) a binding
phenomenon which is incompatible with Fermi liquid behavior was found.
In this regime, the mobility of the $\downarrow$-e$^-$ remains finite.
This non Fermi liquid behavior certainly deserves further investigation.

The author acknowledges stimulating discussions with
E.~M\"uller-Hartmann, Th.~Hanisch, B.~Kleine, and P.~Wurth
as well as with H.~J.~Schulz and A.~Mielke.
It was supported by the SFB 341 of the DFG and by the EEC, grant 
ERBCHRXCT 940438.


\begin{thebibliography}{10}

\bibitem{metzn89a}
W.~Metzner and D.~Vollhardt, Phys. Rev. Lett. {\bf 62}, 324 (1989)

\bibitem{mulle89a}
E.~M\"uller-Hartmann, Z. Phys. B {\bf 74}, 507 (1989)

\bibitem{vollh93}
D.~Vollhardt, in {\em Correlated Electron Systems}, by V.~J. Emery, p.~57
  (World Scientific, Singapore, 1993)

\bibitem{janis91}
V.~Jani\v{s}, Z. Phys. B {\bf 83}, 227 (1991)

\bibitem{georg92ajarre92}
A.~Georges and G.~Kotliar, Phys. Rev. B {\bf 45}, 6479 (1992);
M.~Jarrell, Phys. Rev. Lett. {\bf 69}, 168 (1992)

\bibitem{prusc96georg96}
Th. Pruschke, M.~Jarrell and J.~K. Freericks, Adv. Phys. {\bf 44}, 187(1995);
A.~Georges {\it et al.}, Rev. Mod. Phys. {\bf 68}, 13 (1996)

\bibitem{janis92ajanis92b}
V.~Jani\v{s} and D.~Vollhardt, Int. J. Mod. Phys. {\bf 6}, 731 (1992);
ibid.\ Phys. Rev. B {\bf 46}, 15712 (1992)

\bibitem{donge91donge94bdonge95}
P.~G.~J. van Dongen, Phys. Rev. Lett. {\bf 67}, 757 (1991);
ibid.\ Phys. Rev. B {\bf 50}, 14016 (1994);
ibid.\ Phys. Rev. Lett. {\bf 74}, 182 (1995)

\bibitem{uhrig93buhrig95d}
G.~S. Uhrig and R.~Vlaming, Phys. Rev. Lett. {\bf 71}, 271 (1993);
ibid.\ Ann. Physik {\bf 4}, 778 (1995)

\bibitem{freer93}
J.~K. Freericks, Phys. Rev. B {\bf 47}, 9263 (1993)

\bibitem{freer95}
J.~K. Freericks and M.~Jarrell, Phys. Rev. Lett. {\bf 74}, 186 (1995)

\bibitem{rozen95}
M.~J. Rozenberg, Phys. Rev. B {\bf 52}, 7369 (1995)

\bibitem{held95}
K.~Held, M.~Ulmke and D.~Vollhardt,
Mod. Phys. Lett., {\bf 10}, 203 (1996)

\bibitem{ulmke95}
M.~Ulmke, cond-mat 9512044

\bibitem{hubba63kanam63gutzw63}
J.~Hubbard, Phys. Roy. Soc. {\bf 276}, 238 (1963);
J.~Kanamori, Prog. Theor. Phys. {\bf 30}, 275 (1963);
M.~C. Gutzwiller, Phys. Rev. Lett. {\bf 10}, 159 (1963)

\bibitem{nagao66}
Y.~Nagaoka, Phys. Rev. {\bf 20}, 392 (1966)

\bibitem{mielk91amielk91btasak92mielk93}
A.~Mielke, J. Phys. {\bf A 24}, L73 (1991);
ibid.\ J. Phys. {\bf A24}, 3311 (1991);
H.~Tasaki, Phys. Rev. Lett. {\bf 69}, 1608 (1992);
A.~Mielke and H.~Tasaki, Commun. Math. Phys. {\bf 158}, 341 (1993)

\bibitem{lieb89}
E.~H. Lieb, Phys. Rev. Lett. {\bf 62}, 1201 (1989)

\bibitem{strac93strac94strac95}
R.~Strack and D.~Vollhardt, Phys. Rev. Lett. {\bf 70}, 2637 (1993);
ibid.\ {\bf 72}, 3425 (1994);
ibid.\ J. Low Temp. Phys. {\bf 99}, 385 (1995)

\bibitem{tasak94tasak95}
H.~Tasaki, Phys. Rev. Lett. {\bf 73}, 1158 (1994);
ibid.\ cond-mat 9512169

\bibitem{mulle95}
E.~M\"uller-Hartmann, J. Low Temp. Phys. {\bf 99}, 349 (1995)

\bibitem{fazek90}
P.~Fazekas, B.~Menge and E.~M\"uller-Hartmann, Z. Phys. B {\bf 78}, 69 (1990)

\bibitem{hanis93}
Th. Hanisch and E.~M\"uller-Hartmann, Ann. Physik {\bf 2}, 381 (1993)
and refs.\ therein

\bibitem{wurth95}
P.~Wurth and E.~M\"uller-Hartmann, Ann. Physik {\bf 4}, 144 (1995)

\bibitem{hanis95}
Th. Hanisch {\it et al.}, Ann. Physik {\bf  4}, 303 (1995)

\bibitem{hirsc94}
R.~Hirsch, Thesis Univ.\ K\"oln, (Shaker, Aachen, 1994)

\bibitem{liang94}
S.~Liang and H.~Pang, cond-mat 9404003

\bibitem{wurth96}
P.~Wurth, G.~S. Uhrig and E.~M\"uller-Hartmann,
Ann. Physik, {\bf 5}, 148 (1996)

\bibitem{donge94a}
P.~G.~J. van Dongen and V.~Jani\v{s}, Phys. Rev. Lett. {\bf 72}, 3258 (1994)

\bibitem{fuldec}
P.~Fulde, {\em Electron Correlations in Molecules and
  Solids}, vol. 100 of {\em Solid State Sciences} (Springer, Berlin,
  1993)

\bibitem{petti85}
D.~G. Pettifor and D.~L. Weaire, {\em The Recursion Method and its
Applications}, vol.~58 of
{\em Solid State Sciences} (Springer, Berlin, 1985)

\bibitem{engel95}
J.~R. Engelbrecht and K.~S.~Bedell,
Phys. Rev. Lett. {\bf 74}, 4265 (1995)

\bibitem{polen}
E.~M\"uller-Hartmann, in {\em Proc.\ V.\ Symp.\ Phys.\ of Metals},
by E.~Talik and J.~Szade, p.\ 22 (Poland, 1991)

\bibitem{mullepriv}
E.~M\"uller-Hartmann, unpublished

\bibitem{hanis96}
Th.~Hanisch and G.~S. Uhrig, to be published

\bibitem{uhrig95a}
G.~Uhrig and D.~Vollhardt,
Phys. Rev. B {\bf 52}, 5617 (1995); For
non-hypercubic lattices eq.\ (21) changes slightly.

\end{thebibliography}
\end{document}